\documentclass[conference]{IEEEtran}

\usepackage{acro}
\usepackage{microtype}
\usepackage{xcolor}
\usepackage{comment}
\usepackage{etoolbox} %
\usepackage{verbatim} %
\usepackage{xcolor-material} %
\usepackage{xstring}
\usepackage{adjustbox}
\usepackage{booktabs}
\usepackage{siunitx}
\usepackage{rotating}
\usepackage{graphicx}
\usepackage{makecell}
\usepackage{url}

\newbool{showoutline}
\newbool{feedback-highlight}

\setbool{showoutline}{true}
\setbool{feedback-highlight}{false}

\newcommand{\update}[1]{\ifbool{feedback-highlight}{{\color{MaterialPink}#1}}{#1}}

\usepackage{xparse}
\NewDocumentEnvironment{outline}{o}
{
\IfValueTF {#1}{
\comment
}{}
\small\color{MaterialGreen}\setstretch{.8}
\setlist[itemize]{noitemsep, topsep=0pt}}
{
\IfValueTF {#1}{
\endcomment
}{}
}

\ifbool{showoutline}{}{
\AtBeginEnvironment{outline}{\comment}%
\AtEndEnvironment{outline}{\endcomment}
}
\usepackage{xspace}

\usepackage{etoolbox}
\preto\chapter\acresetall

\usepackage{mathtools}

\usepackage{csquotes}

\colorlet{punct}{red!60!black}
\definecolor{delim}{RGB}{20,105,176}
\colorlet{numb}{magenta!60!black}
\definecolor{codegreen}{rgb}{0,0.6,0}
\definecolor{codegray}{rgb}{0.5,0.5,0.5}
\definecolor{codepurple}{rgb}{0.58,0,0.82}
\definecolor{backcolour}{rgb}{0.95,0.95,0.92}
\usepackage{multirow}

\usepackage{afterpage}

\DeclareSIUnit\transfer{T}

\usepackage{hhline}

\usepackage{tikz}
\usepackage{tikzscale}
\usepackage{pgf}

\newcommand*\emptycirc[1][1ex]{\tikz\draw[thick] (0,0) circle (#1);} 
\newcommand*\halfcirc[1][1ex]{%
  \begin{tikzpicture}
  \draw[fill] (0,0)-- (90:#1) arc (90:270:#1) -- cycle ;
  \draw[thick] (0,0) circle (#1);
  \end{tikzpicture}}
\newcommand*\fullcirc[1][1ex]{\tikz\filldraw[thick] (0,0) circle (#1);} 

\definecolor{simulation}{HTML}{f1a340}
\definecolor{fischertechnik}{HTML}{998ec3}
\DeclareAcronym{TCP}{short=TCP, long=Transmission Control Protocol}

\DeclareAcronym{ICS}{short=ICS, long = Industrial Control System}
\DeclareAcronym{PLC}{short=PLC, long = Programmable Logic Controller}
\DeclareAcronym{SCADA}{short=SCADA, long=Supervisory Control and Data Acquisition}
\DeclareAcronym{IT}{short=IT, long=Information Technology}
\DeclareAcronym{OT}{short=OT, long=Operational Techology
}
\DeclareAcronym{HMI}{short=HMI, long=Human-Machine Interface}
\DeclareAcronym{DMZ}{short=DMZ, long= demiliterized zone}

\DeclareAcronym{DES}{short=DES, long=Discrete Event Simulation}
\DeclareAcronym{VC}{short=VG, long=vacuum gripper}
\DeclareAcronym{MPU}{short=MPU, long=multi-processing unit}
\DeclareAcronym{IDS}{short=IDS, long=Intrusion Detection System}
\DeclareAcronym{IIDS}{short=IIDS, long= Industrial IDS}
\DeclareAcronym{SDR}{short=SDR, long=Software-Defined Radio}
\DeclareAcronym{IAT}{short=IaT, long=Inter-Arrival Time}
\DeclareAcronym{UE}{short=UE, long=User Equipment}
\usepackage{enumitem}

\usepackage{flushend}
\usepackage{cite}
\usepackage{amsmath,amssymb,amsfonts}
\usepackage{algorithmic}
\usepackage{graphicx}
\usepackage{textcomp}
\usepackage{xcolor}

\usepackage{tcolorbox} 
\newtcolorbox{takeaway}{
  left=0.5mm,
  right=0.5mm,
  bottom=0.5mm,
  top=0.5mm,
  before upper={\textit{Take Away: }}
}

\begin{document}

\makeatletter
\def\ps@IEEEtitlepagestyle{%
\def\@oddfoot{\parbox{\textwidth}{\footnotesize
Author's version of a paper accepted for publication in Proceedings of the 2025 IEEE 50th Conference on Local Computer Networks (LCN).
\\
\textcopyright{} 2025 IEEE.
Personal use of this material is permitted.
Permission from IEEE must be obtained for all other uses, in any current or future media, including reprinting/republishing this material for advertising or promotional purposes, creating new collective works, for resale or redistribution to servers or lists, or reuse of any copyrighted component of this work in other works.\vspace{1.2em}}
}%
}
\makeatother

\newcommand{\name}{CoFacS}

\title{\name \ --  Simulating a Complete Factory to Study the Security of Interconnected Production}

\author{\IEEEauthorblockN{Anonymous Author(s)}}

\author{\IEEEauthorblockN{%
Stefan Lenz\IEEEauthorrefmark{1}, 
David Schachtschneider\IEEEauthorrefmark{1}, 
Simon Jonas\IEEEauthorrefmark{1},
Liam Tirpitz\IEEEauthorrefmark{3},
Sandra Geisler\IEEEauthorrefmark{3},
Martin Henze\IEEEauthorrefmark{1}\IEEEauthorrefmark{4}
}
\IEEEauthorblockA{%
\IEEEauthorrefmark{1}\textit{Security and Privacy in Industrial Cooperation}, RWTH Aachen University, Germany \\
\IEEEauthorrefmark{3}\textit{Data Stream Management and Analysis}, RWTH Aachen University, Germany \\
\IEEEauthorrefmark{4}\textit{Cyber Analysis \& Defense}, Fraunhofer FKIE, Germany\\
\{lenz, schachtschneider, jonas, henze\}@spice.rwth-aachen.de \(\cdot\)  \{tirpitz, geisler\}@dbis.rwth-aachen.de
}
}

\maketitle

\begin{abstract}
While the digitization of industrial factories provides tremendous improvements for the production of goods, it also renders such systems vulnerable to serious cyber-attacks.
To research, test, and validate security measures protecting industrial networks against such cyber-attacks, the security community relies on testbeds to simulate industrial systems, as utilizing live systems endangers costly components or even human life.
However, existing testbeds focus on individual parts of typically complex production lines in industrial factories.
Consequently, the impact of cyber-attacks on industrial networks as well as the effectiveness of countermeasures cannot be evaluated in an end-to-end manner.
To address this issue and facilitate research on novel security mechanisms, we present \name{}, the first COmplete FACtory Simulation that replicates an entire production line and affords the integration of real-life industrial applications. 
To showcase that \name \ accurately captures real-world behavior, we validate it against a physical model factory widely used in security research.
We show that \name\ has a maximum deviation of 0.11\% to the  physical  reference, which enables us to study the impact of physical attacks or network-based cyber-attacks.
Moreover, we highlight how \name\ enables security research through two cases studies surrounding attack detection and the resilience of 5G-based industrial communication against jamming.
\end{abstract}
\IEEEpeerreviewmaketitle


\section{Introduction}

The digitization of factories is the foundation for efficient and adaptable production of goods~\cite{pennekamp2019towards}.
However, the resulting increase in connectivity exposes complex industrial control systems with numerous interacting components to cyberattacks~\cite{knapp2024_CH3}.
Attacks targeting factories and production lines can, e.g., halt the production, damage  components, or even physically harm workers, posing a great risk to companies around the globe.

However, researching, testing, and evaluating novel methods to secure factories and production lines cannot be performed on actual live systems, due to strict availability and safety requirements~\cite{conti2021}.
Therefore, the security community has to rely on testbeds~\cite{conti2021}.
Although \textit{physical testbeds} (e.g., \cite{morris2011}) provide the highest degree of realism, they are often hard to access, inflexible, and expensive to build~\cite{conti2021}. %
Moreover, it is impractical to examine attacks that might irreversibly damage expensive equipment of the physical testbed.
In contrast, \textit{virtual testbeds} provide a high degree of flexibility and accessibility, but must be validated against a real-life system to provide sufficiently realistic results.
As recent surveys show~\cite{conti2021}, there exist a variety of virtual testbeds of industrial factories for security research (e.g., ~\cite{wattson, grfics,dehlaghi-ghadim2023}).
However, all publicly available testbeds that specifically focus on factories capture only a single step of typically involved production lines~\cite{DVCP,xie2018_VTET,grfics, dehlaghi-ghadim2023}.
Thus, these testbeds do not adequately capture the complexity of modern factories, including the interaction of different production systems, such as the transferal from one production step to the next, their timings, or additional physical properties. %

To address this gap and thus lay the foundation for comprehensively studying the security of factories, we propose \textit{\name}; a simulation of an industrial factory that covers a complete production line, from arrival of raw material in the factory to the completion of  the finished product.
To realistically and comprehensively cover the properties of a real factory, \name{} not only accurately captures the underlying physical processes but also simulates all control components such as \acp{PLC}, a \ac{SCADA} system, as well as the corresponding industrial network.

To ensure accurate behavior of all testbed components, we provide a comprehensive validation of \name{}.
We utilize the Fischertechnik Learning Factory 4.0~\cite{FTfactory} as \textit{physical} reference, which provides a lab-scale replica of a complete production process. 
We choose this reference model, since it is already successfully utilized in other research~\cite{licster2019, Malburg2022, sala2023developmentDigitalShadow, alsabbagh2024} and generally available to other researchers for reproducibility.

\noindent \textbf{Contributions.} 
With the goal to enable research, testing, and evaluation of novel mechanisms to strengthen the security of industrial factories, we present the following contributions.

\begin{enumerate}[topsep=0em,leftmargin=1.5em] 
    \item We provide \name{}, a comprehensive, freely-available, virtual testbed of an industrial factory, simulating physical processes, \ac{PLC} logic, network communication, and SCADA application of a complete production line (Sec.~\ref{sec:design}).
    \item We validate \name{}'s accuracy against the Fischertechnik Learning Factory 4.0~\cite{FTfactory}, a realistic physical factory simulation often used in security research (Sec.~\ref{sec:validation}).
    \item We show the usability of \name \ for security research by evaluating the behavior of the simulation and the physical reference model under different attack scenarios (Sec.~\ref{sec:attack}). 
    \item We further exemplify \name{}'s versatility as an enabler for security research, by studying the resilience of 5G-based production system
    a wireless production system utilizing a real 5G channel and \name{}'s capability to research attack detection approaches (Sec.~\ref{sec:intrusion}).
\end{enumerate}

\noindent \textbf{Availability Statement.} 
As a novel virtual factory, we provide full access to the code base of \name{}\footnote{We provide code and results under \textit{\url{github.com/RWTH-SPICe/CoFacS}}} and the exemplary attacks, encouraging researchers to use our testbed for their own security research.
To enable reproducibility, we also make all results underlying our evaluation available.  

\section{The Need for Complete Factory Simulation}

To motivate the need for complete factory simulation covering all steps of a production process, we first provide a short introduction to \acp{ICS} and testbeds (Sec.~\ref{sec:background}).
Subsequently, we analyze the current landscape of testbeds, highlighting the gap in existing virtual testbeds (Sec.~\ref{sec:related_work}).

\subsection{Industrial Control Systems \& Testbeds}
\label{sec:background}

\acfp{ICS} monitor and control complex physical processes, e.g., in factories, and thus serve as the backbone of digitized production. 
The components that make up an ICS are typically grouped into field bus, control, and supervisory level~\cite{purdueModel1994}. %

The \textit{field bus level} comprises all devices directly interacting with the physical process, such as sensors and actuators. 
In the execution of a control loop (i.e., one cycle of the \ac{ICS} logic), these devices forward the process state to the \textit{control level} and in turn receive commands to interact with the physical process.
To this end, controllers (e.g., \acp{PLC}) respond to the current process state as captured by sensors according to their control logic and send commands back to actuators.
Complex manufacturing processes necessitate the use of multiple sensors and actuators, each demanding prompt responses to ensure the safe control of the underlying physical process.
Consequently, \acp{ICS} are composed of several \acp{PLC}, with each PLC responsible for controlling a specific part of the system, such as an individual manufacturing step.
Additionally, the controllers send updates to devices on the \textit{supervisory level}.
These supervisory devices include, e.g., a \ac{SCADA} system, to facilitate direct human interaction to monitor the state of the physical process and perform higher-level control such as which parts to produce.

Since such \acp{ICS} which control highly critical processes have strong availability and safety requirements~\cite{conti2021}, testing of novel security measures cannot be performed on live systems.
Thus, researchers rely on replicas of \acp{ICS}, which represent industrial processes in physical, virtual (e.g., a digital simulation), or hybrid testbeds as a combination of the two~\cite{conti2021}.
Although \textit{physical} testbeds allow the most realistic representation of \acp{ICS}, they suffer from high costs and inflexibility, due to set up times, which also cause scalability issues.
Additionally, physical testbeds often provide high barriers for accessibility. %
In contrast, \textit{virtual} testbeds provide high accessibility and flexibility, since researchers only need means to execute the testbed (e.g., a simulation) and experiments can be done quickly.
However, this abstraction entails the disadvantage of possibly providing less accurate results.
Thus, researchers must thoroughly validate virtual testbeds to ensure correct results.
Then, validated virtual testbeds allow testing of additional scenarios that would, e.g., be too dangerous to perform in a physical testbed~\cite{conti2021}.

Consequently, validated virtual testbeds are crucial to enable various security research.
For example, simulations enable studying trends of modern industry such as wireless 5G communication.
As jamming attacks potentially cause serious damage to physical components (e.g., by rendering the \ac{ICS} unable to react to emergency scenarios), virtual testbeds allow their execution in a safe environment.
Thus, they facilitate the research of new security-trends for safety-reliant production systems.
Likewise, validated virtual testbeds with their ability to gather data such as network traffic captures, enable the study of further defense mechanisms such as \acp{IDS}.
These systems utilize the predictability of \acp{ICS} to detect attacks in communication patterns (e.g., timings~\cite{lin2018iat}, packet sequences~\cite{ferling2018DTMC}) or the physical process (e.g, \cite{wolsing2022simple}).

\begin{takeaway}
    \acp{ICS} serve as the backbone of modern digitized production.
    Due to availability and safety requirements, testing security mechanisms for \acp{ICS} is challenging on live-systems.
    Thus, researchers utilize testbeds to gain insight into such systems. 
    As virtual testbeds provide the highest degree of flexibility and accessibility without damaging physical components, they are a good solution to gain deep insights into \ac{ICS}. 
\end{takeaway}

\subsection{Related Work on ICS Testbeds}
\label{sec:related_work}

Given these benefits of (virtual) testbeds for \ac{ICS} security research, we now analyze already available testbeds.  
As a recent survey comprehensively summarizes the landscape of \ac{ICS} testbeds~\cite{conti2021}, we specifically focus on research leveraging the Fischertechnik Learning Factory 4.0~\cite{FTfactory} and virtual testbeds related to our approach.
We summarize these works in Table~\ref{tb:related-work} based on their \textit{scenario}, the support of \textit{network}, \textit{physical}, \textit{control logic}, and \textit{\ac{SCADA}} simulation, their \textit{accessibility}, and whether they realize a \textit{complete factory simulation}.

    \begin{table}[t]
        \tiny
        \centering
        \caption{The current landscape of hybrid and virtual ICS testbeds for interconnected production includes a variety of different approaches. We compare (1) scenario, (2) completeness as a factory simulation, (3) network emulation, (4) physical simulation, (5) SCADA simulation, (6) control logic simulation, and (7) the accessibility, highlighting the need for a complete virtual testbed to enable comprehensive security research for interconnected production.
        }
        \label{tb:related-work}
        \begin{tabular}{l l l c c c c c c}\toprule
            \makecell{\rotatebox[origin=c]{90}{\textbf{Type}}} &
            \makecell{\rotatebox[origin=l]{0}{\textbf{Name}}} &
            \makecell{\rotatebox[origin=l]{0}{\textbf{Scenario}}} &
            \makecell{\rotatebox[origin=c]{90}{\parbox{1cm}{\textbf{Complete Factory \\ Simulation}}}} & 
            \makecell{\rotatebox[origin=c]{90}{\parbox{1cm}{\textbf{Network \\ Emulation}}}} & 
            \makecell{\rotatebox[origin=c]{90}{\parbox{1cm}{\textbf{Physical \\ Simulation}}}} & 
            \makecell{\rotatebox[origin=c]{90}{\parbox{1cm}{\textbf{SCADA \\ Simulation}}}} &
            \makecell{\rotatebox[origin=c]{90}{\parbox{1cm}{\textbf{Control Logic \\ Simulation}}}} & 
            \makecell{\rotatebox[origin=c]{90}{\textbf{Accessibility}}} \\ \midrule
            hybr. &LICSTER~\cite{licster2019} & Factory & \halfcirc & \fullcirc & \emptycirc & \fullcirc & \fullcirc &\fullcirc \\
            hybr. &Foley et al.~\cite{foley2018} & Factory & {\footnotesize ?} & \fullcirc & \emptycirc & \halfcirc & \emptycirc & \fullcirc \\ 
            phys. & Gardiner et al.~\cite{gardiner2019} & Factory & \halfcirc & \halfcirc & \halfcirc  & \fullcirc & \emptycirc & \emptycirc \\ \midrule
            \multirow{14}{*}{\rotatebox{90}{virtual}}&Davis et al.~\cite{davis2006} & Power Grid & {\scriptsize ---} & \fullcirc & \fullcirc & \fullcirc & \fullcirc & \fullcirc \\
            &Koganti et al.~\cite{koganti2017} & Power Grid & {\scriptsize ---} & \emptycirc & \emptycirc & \fullcirc & \fullcirc & \emptycirc \\
            &RICS-el~\cite{almgren2019} & Power Grid & {\scriptsize ---} & \fullcirc & \fullcirc & \fullcirc & \fullcirc & \emptycirc \\
            &Singh et al.~\cite{singh2015} & Power Grid & {\scriptsize ---} &  {\footnotesize ?}  & \fullcirc & \fullcirc & \fullcirc & \emptycirc \\
            &TasSCS~\cite{tasSCS} & Power Grid & {\scriptsize ---} & \fullcirc & \fullcirc & \fullcirc & \fullcirc & \emptycirc \\
            &Wattson~\cite{wattson} & Power Grid & {\scriptsize ---} &  \fullcirc & \fullcirc & \fullcirc & \fullcirc & \fullcirc \\
            &Alves et al.~\cite{alves2016} & Gas Pipeline & {\scriptsize ---}  & \fullcirc & \fullcirc & \fullcirc & \fullcirc & \emptycirc \\
            &Morris et al.~\cite{morris2015industrial} & Gas Pipeline & {\scriptsize ---} & \fullcirc & \fullcirc & \fullcirc & \fullcirc & \emptycirc \\
            &SCADAVT-A~\cite{SCADAVT-A} & Water Pipeline & {\scriptsize ---} &  \fullcirc & \fullcirc & \fullcirc & \fullcirc & \emptycirc \\
            &DVCP~\cite{DVCP} & Chemical Plant& \emptycirc & \halfcirc & \fullcirc & \fullcirc & \fullcirc & \fullcirc \\
            &VTET~\cite{xie2018_VTET} & Chemical Plant & \emptycirc & \fullcirc & \fullcirc & \fullcirc & \fullcirc & \emptycirc \\ 
            &GRIFCS~\cite{grfics} & Chemical Plant & \emptycirc & \fullcirc & \fullcirc & \fullcirc & \fullcirc & \fullcirc \\
            &ICSSIM~\cite{dehlaghi-ghadim2023} & Factory & \emptycirc & \fullcirc  & \fullcirc & \fullcirc & \fullcirc & \fullcirc \\ 
            &Sala et al.~\cite{sala2023developmentDigitalShadow} & Factory & \halfcirc &  {\footnotesize ?}  & \fullcirc & \fullcirc & \fullcirc & \emptycirc \\ \cmidrule{2-9}
            &\textit{\name{} (this paper)} & \textit{Factory} & \fullcirc & \fullcirc & \fullcirc & \fullcirc & \fullcirc & \fullcirc \\ \bottomrule
        \end{tabular}
        {\tiny \raisebox{-.6ex}{{\tiny \emptycirc}}\,: No support/access \hfil  \raisebox{-.6ex}{{\tiny \halfcirc}}\,: Supported, but not provided/access limited \hfil  \raisebox{-.6ex}{{\tiny \fullcirc}}\,: Provided/full access \hfil  \raisebox{-.3ex}{{\footnotesize ?}}:  Unclear}
    \end{table}

\noindent \textbf{Physical and Hybrid Testbeds.} 
We discuss physical and hybrid testbeds utilizing the Fischertechnik hardware, showing its viability for researching modern production environments.
LICSTER~\cite{licster2019} is a hybrid testbed to simulate control logic and \ac{SCADA} application based on individual Fischertechnik components.
Gardiner et al.~\cite{gardiner2019} present a physical \ac{ICS} testbed including the Fischertechnik model factory~\cite{FTfactory} to research guidelines for building testbeds.
They confirm the usefulness of the Fischertechnik factory for setting up new testbeds due to high fidelity.
Similarly, Foley et al.~\cite{foley2018} apply the Fischertechnik factory to building a hybrid model of a complete production line.
Although these works show the benefits of the Fischertechnik factory~\cite{FTfactory} for security research, they all require a physical model factory, limiting accessibility and flexibility due to costs and potential damage to physical components.

\noindent \textbf{Virtual Testbeds.} 
Focussing on virtual testbeds for \ac{ICS} and specifically production lines, we observe a variety of research.

Research on virtual testbeds focuses on simulations to safely research the impact of cyber-attacks in \emph{power grids}~\cite{davis2006,koganti2017,almgren2019,singh2015,tasSCS,wattson}.
While the physical process of power grids is clearly distinct from modern multi-step production lines in factories, those works highlight important aspects of virtual testbeds: 
authentically capturing the physical process~\cite{singh2015}, replicating realistic traffic patterns and network architectures~\cite{koganti2017,tasSCS}, fully capturing \ac{ICS} assets~\cite{almgren2019}, and thoroughly validating against a physical reference~\cite{wattson}.
To research \emph{gas pipeline} security, two virtual testbeds replicate lab-scale physical models to create authentic representations~\cite{alves2016,morris2011} 
demonstrating the advantages w.r.t.\ validation.
Likewise, a virtual testbed simulating a \emph{water pipeline}~\cite{SCADAVT-A} highlights the need for virtual simulations to safely research attack scenarios with potentially devastating consequences.
Similarly, simulations of a \emph{chemical plant}~\cite{DVCP,xie2018_VTET,grfics} highlight this necessity with catastrophic outcomes such as an exploding reactor.

Moving towards \emph{factory} simulation, ICSSIM~\cite{dehlaghi-ghadim2023} replicates a bottle filling plant to simulate a ``robust'' process that can quickly recover from a fault state enabling researchers to test several attack scenarios without waiting for the process to recover. 
However, by focussing on only one process step, ICSSIM misses the interaction between different components.
Finally, to test the ability of students to create a digital shadow of a complex production line, Sala et al.~\cite{sala2023developmentDigitalShadow} utilize the Fischertechnik Learning Factory 4.0~\cite{FTfactory}. 
Not focussing on security, they neither create an authentic simulation of a complete factory including network communication nor evaluate the authenticity and accuracy of their simulation.
Still, they show the general suitability of the Fischertechnik factory as a baseline to create a complete factory simulation.

\begin{takeaway}
    There is no virtual \ac{ICS} testbed covering a complete production process \emph{and} fully replicating all aspects of modern digitized production (i.e., physical process, control logic, \ac{SCADA}, and network).
    To address this gap, we propose \name{}, a \emph{complete} virtual factory to facilitate the study of interconnected production.
\end{takeaway}

\section{The \name \ Complete Factory Simulation}
\label{sec:design}

\begin{figure}[t]
    \centering
    \includegraphics[scale=0.35]{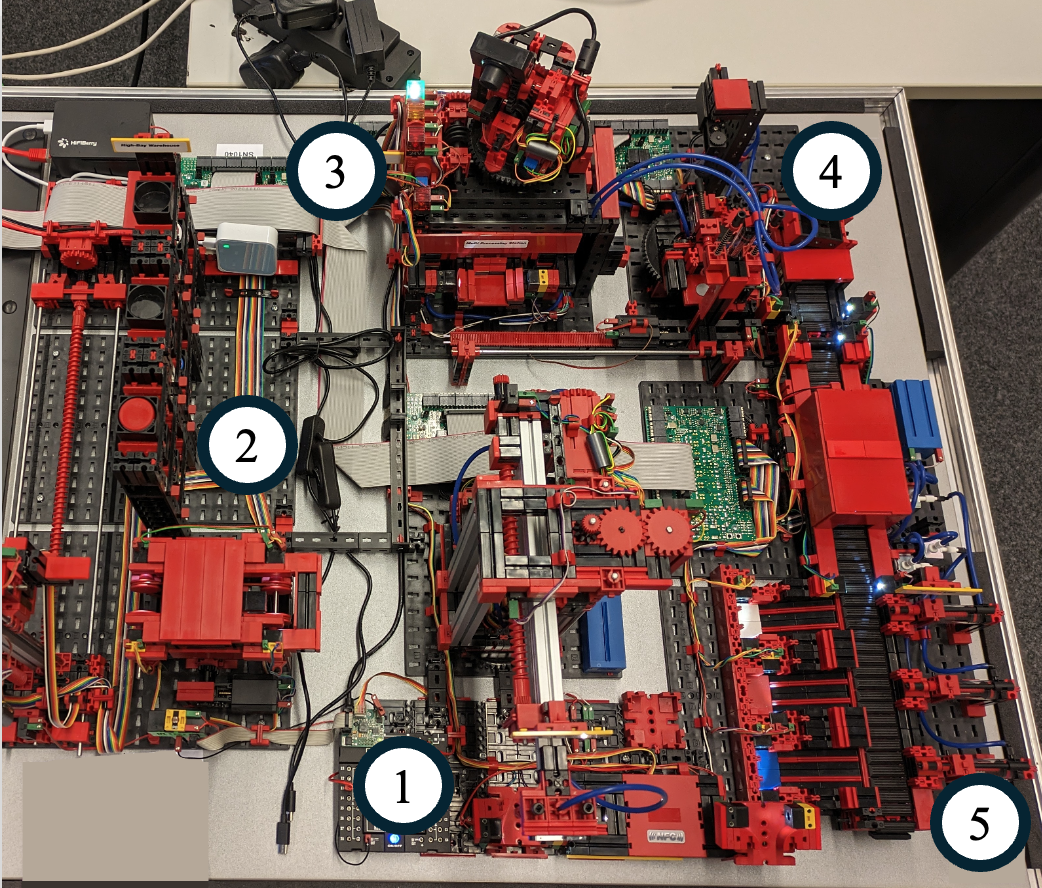}
    \caption{The Fischertechnik Learning Factory 4.0 is a lab-scale replica of a complete production line covering (1) a vacuum gripper, (2) a high-bay warehouse, (3) a furnace, (4) a mill, and (5) a sorting station. It serves as a widely-used reference architecture for (security) research~\cite{licster2019,gardiner2019,foley2018, sala2023developmentDigitalShadow}.}
    \label{fig:factory}
\end{figure}

To develop, test, and evaluate novel security measures for interconnected production,
we propose \name, the first simulation of a complete factory.
To facilitate accessibility and reproducibility~\cite{uetz2021socbed}, as well as to validate our simulation, we utilize the Fischertechnik Learning Factory 4.0~\cite{FTfactory} as a physical reference.
This factory, as shown in Fig.~\ref{fig:factory}, encompasses a complete production process from the delivery of raw material to completion of the product, including process control logic, network communication, and a \ac{SCADA} system as a table-sized physical model.
Additionally, the Fischertechnik factory can be bought as a ready-to-use package or as individual components~\cite{licster2019} without requiring intensive setup.
Therefore, we choose this model as a physical reference for our simulation, enabling reproducibility but also flexibility to place individual, physical components in a hybrid factory testbed.

As a virtual model, \name \ provides an accurate simulation of all physical components, the process logic, and \ac{SCADA} monitoring of the factory. 
To serve its purpose as a security testbed, \name \ also precisely emulates the industrial network, allowing for the integration for real applications and attack tools in this network.
Furthermore, \name \ allows extracting artifacts for further analysis such as the sensor and actuator states or recordings of the network traffic.

To introduce our testbed, we first describe the individual components and the production process of the Fischertechnik Learning Factory 4.0 reference factory (Sec.~\ref{sec:overview}), before we detail the technical realization of \name \ (Sec.~\ref{sec:implementation}).

\subsection{The Simulated Physical Production Line}
\label{sec:overview}

The modeled process of the Fischertechnik Learning Factory 4.0~\cite{FTfactory} consists of five different components: a \textit{\acf{VC}}, a \textit{high-bay warehouse}, a \textit{\ac{MPU}}, comprising a \textit{furnace} and a \textit{mill}, and a \textit{sorting station} (Fig.~\ref{fig:factory}).
The process itself mimics a complete production line by using small ``cylinders'' (i.e., miniature hockey pucks) in red, white, or blue as the material and end-product. 
These cylinders then pass each component, replicating the arrival, storage, manufacturing and finally sorting of a product.

\noindent
\textbf{Step 1 -- Material Input \& Vacuum Gripper.} %
Once a new (``raw'') cylinder arrives at factory, the \textit{\ac{VC}} transports the material to the high-bay warehouse to store until needed to complete an order.
To do so, the \textit{\ac{VC}} uses its arm with a suction mechanism to pick up new cylinders  (Fig.~\ref{fig:factory}-{1}).
The \ac{VC} senses the rotation angle, horizontal, and vertical position of the rotating arm, and transmits these values to the \ac{PLC}.
The \ac{PLC} controls the electric motor of the \ac{VC}, to reach pre-defined destinations (e.g., delivery and pickup station).

\noindent
\textbf{Step 2 -- High-Bay Warehouse.} %
The \textit{high-bay warehouse} (Fig.~\ref{fig:factory}-2) stores raw material (i.e., recently delivered cylinders) in a high-bay rack until they are needed for the production process (i.e., ordered through the \ac{SCADA} application).
To this end, a cantilever picks up a cylinder at the drop-off location of the \ac{VC} and transports it to a free spot in the high-bay rack. %
The factory keeps track of each stored cylinder using a triple of X and Y coordinates and the color of the cylinder.  
Once a product order arrives through the \ac{SCADA} system, the warehouse ``unloads'' a cylinder in the requested color by performing these actions in reverse order, placing the cylinder at the \ac{VC}'s pick-up location. 

\noindent
\textbf{Step 3 \& 4 -- Multi-Processing Unit.} %
The \acf{MPU} replicates the production steps of a production line with two parts, a \textit{furnace} and a \textit{mill}, producing a ``finished'' product ready for delivery.
The first part of the \ac{MPU}, the \textit{furnace} (Fig.~\ref{fig:factory}-3), bakes the raw cylinder.
To this end, the \ac{VC} transports a cylinder from the warehouse to a platform in front of the furnace.
Then, the platform transports the cylinder into the chamber of the furnace, where the cylinder resides for a per-order customizable period of time. 
The duration of the firing process can be customized from the \ac{SCADA} interface.
To indicate the firing process of a cylinder, the furnace controls an LED light that is activated for the particular duration.
Once this process is complete, the cylinder leaves the firing chamber and the \ac{MPU} transfers the cylinder to the \textit{mill} (Fig.~\ref{fig:factory}-4), the second and final step of the production process. 
This step simulates finishing the product in, e.g., a CNC machine.
To replicate this behavior, the physical reference factory actuates an electrical motor that turns a gear wheel above the current cylinder.
Finally, a piston pushes the now finished cylinder onto a conveyor of the last component, the \textit{sorting station}.

\noindent
\textbf{Step 5 -- Sorting \& Delivery.} %
After being processed in the \ac{MPU}, the finished cylinder arrives on the conveyor belt of the \textit{sorting station} (Fig.~\ref{fig:factory}-5).
The sorting stations purpose is to organize finished cylinders based on their colors for the delivery.
Thus, the cylinder first passes through a light-barrier, which activates a door to the color-sensing enclosure.
Within this light-proof enclosure, an LED creates a bright flash.
Then, a sensor compares the reflection of the cylinder to the baseline thus receiving a distinct value for each of the three colors (i.e, red, white, or blue).
Following the reading, the cylinder passes through another light-barrier, this time activating a timer.
Finally, based on the timer, a piston activates to sort the cylinders into bays, with each bay designated for a specific color, thus completing the production process.

\begin{takeaway}
    The considered production process consists of five distinct components, each with an own purpose, requiring different control logic, sensors, actors, and timings, adaptable to specific demands of individual orders.
    Additionally, to produce a finished product, the components must correctly work together.
    To be able to capture the interplay of these components, \name{} has to thoroughly simulate the entire production process. 
\end{takeaway}

\subsection{Realizing a Complete Factory Simulation}
\label{sec:implementation}

\begin{figure}[t]
    \centering
    \includegraphics[width=\columnwidth]{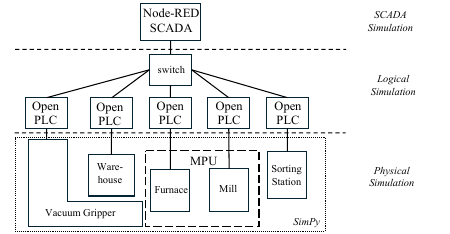}
    \caption{\name \ is a complete factory simulation replicating physical components, \ac{PLC} control logic with IEC-61131-3 compliant OpenPLC code, a fully usable Node-RED \ac{SCADA} application, and an authentic network emulation.} %
    \label{fig:system_overview}
\end{figure}

To build \name \ and correctly capture the complexity of the physical reference model, we partition \name \ into four components: the \textit{physical}, \textit{logical}, and \textit{\ac{SCADA} simulation}, and \textit{network emulation}, which we introduce in the following.
Fig.~\ref{fig:system_overview} visualizes the structure, all components, and the network configuration of \name{}. 

\noindent
\textbf{Physical Simulation.}
The physical simulation replicates the physical behavior of the five components (i.e., \ac{VC}, warehouse, furnace, mill, and sorting station) as well as the cylinders.
To this end, we utilize \ac{DES} to represent the physical state of the factory.
Assuming that the current state directly results from a previous state, \ac{DES} simulates time as discrete steps where an event occurs. 
We base our choice on the fact that \acp{PLC} also operate on polling cycles and thus control the physical state of the process in fixed time intervals, which fits the simulation behavior of \ac{DES} nicely.
Furthermore, this choice enables \name \ to save computational resources, allowing execution in light-weight environments.%

More specifically, we simulate the physical properties of the production line components using SimPy~\cite{simpy}, a Python-based \ac{DES} framework, capturing all necessary variables to execute each component.
We create one process for each component within SimPy's environment to recreate the respective physical behavior.
To keep track of the cylinders that are present in the factory, each process monitors a list of which cylinders to handle.
Additionally, processes may trigger event for other components, such as the \ac{VC} triggering a light barrier.
We recreate the \ac{VC}'s movement using three variables, for the horizontal, vertical positions, and for the rotation angle.  
Additionally, we represent the cylinders as files within \name{}'s codebase, which eases their creation and deletion, thus enabling physical attacks, such as forcibly removing cylinders during production.
To parameterize our simulation, we recorded the behavior of the physical testbed during complete production cycles, fixing color, storage position, and processing times.

\noindent
\textbf{Logical Simulation.}
The logical simulation layer, i.e., the simulation of the control logic, consists of five \acp{PLC}, one for each of the production components.
With this design choice, we allow for adjustments in the control logic, whether simulating benign adaptation of the process or acting as an attacker.
Additionally, it enables scalability of the simulation, by allowing the addition of more components.
We implement control programs for all five \acp{PLC} in the IEC-61131-3 compliant OpenPLC framework~\cite{alves2014openplc} and the structured text programming language.
Furthermore, OpenPLC enables users to exchange the PLC logic with any IEC-61131-3 compliant code allowing for testing of real-world logic in a safe environment.
To accurately control the physical hardware, we set the default \ac{PLC} cycle time to that of the physical reference (i.e., \SI{20}{ms}).
Additionally, \name{} includes OpenPLC's web-interface, which enables adaptation of the \ac{PLC} logic during live operation of the virtual factory.

\noindent
\textbf{SCADA Simulation.}
The \ac{SCADA} application performs higher-level supervision and control of our virtual factory by forwarding order-specific parameters (e.g., the firing time of an individual cylinder) to the \acp{PLC}.
To set such parameters, users can directly interact with the \ac{SCADA} application, ordering new products, adapting the workflow, or monitoring the status of the factory.
We implement this application using Node-RED~\cite{nodered}, an open-source \ac{SCADA} system, which is also used in real-life industrial applications and in the Fischertechnik Learning Factory 4.0~\cite{FTfactory}.
Furthermore, Node-RED provides a web-interface enabling detailed real-time monitoring, interactive ordering, and automation of the virtual factory.

\noindent
\textbf{Network Emulation.}
As a part of an accurate depiction of a complete factory, \name{} provides an emulation of the network communication.
We emulate  connections between the \ac{SCADA} and the \acp{PLC} (connected via a network switch), and between the \acp{PLC} and the physical simulation of their respective production line components (Fig.~\ref{fig:system_overview}).
To achieve authentic traffic patterns, we utilize ModbusTCP~\cite{modbustcp}, the most commonly used communication protocol for industrial networks~\cite{conti2021}. 
To emulate the communication channel, we use Containernet~\cite{containernet}, a ``containerized'' fork of the Mininet~\cite{mininet} simulator, which enables the emulation of crucial properties for industrial deployments, e.g., latency~\cite{hiller2018secure} or bandwidth~\cite{henze2016cppl}, which we utilize to emulate physical distance between the \ac{SCADA} and \ac{PLC} layers.
Additionally, using containers enables the flexible integration of additional industrial components and applications if needed.
Furthermore, this choice allows replacing the emulated network with real-world network components by deploying each container on a different physical host.
We exemplify how this can be utilized in Section \ref{sec:use-cases:5g}.

\begin{takeaway} 
    We present \name{} a \textit{complete} virtual factory, consisting of all components (i.e., physical process, control logic and \ac{SCADA} monitoring) also present in modern interconnected production lines.
    To achieve high realism, \name{} utilizes industrial protocols and frameworks as found in real-world deployments.
\end{takeaway}

\section{Testbed Accuracy \& Resource Requirements}
\label{sec:validation}

To verify the accuracy of \name, we compare its behavior to a physical Fischertechnik Learning Factory 4.0~\cite{FTfactory}.
Concretely, we compare all components during a complete production process and study the \ac{VC} as the most complex component of the factory in more detail. %
We evaluate on fresh data collected \emph{after} \name{} has been parameterized to avoid overfitting.

\noindent
\textbf{Evaluation Setup.}
To demonstrate that \name{} is a lightweight testbed, we conduct all experiments on a resource-constrained Raspberry Pi 5 equipped with a \SI{2.4}{GHz} quad-core ARM CPU and \SI{4}{GB} of RAM. We chose this hardware due to its low cost and accessibility, making \name{} available to researchers and students worldwide without requiring powerful and expensive equipment. Furthermore, the limited resources provide an ideal foundation for assessing the resource requirements of our simulation.

\begin{figure}
  \includegraphics{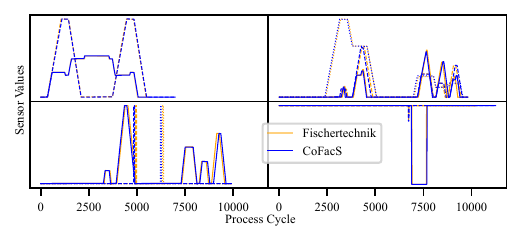}
  \caption{We showcase the correctness of \name{} (\textcolor{blue}{blue}) by comparing the behavior during a complete production cycle with the Fischertechnik factory (\textcolor{orange}{orange}). 
  The sensor values (different line styles) of all components, i.e., the warehouse (upper left), the \ac{VC} (upper right), the \ac{MPU} (lower left), and the sorting station (lower right), substantially overlap during the process execution, showing correct behavior of \name{}.}
  \label{fig:example}
\end{figure}

\noindent
\textbf{Example Production Process.}
We demonstrate \name's correct behavior by conducting an example production process.
We simulate the arrival of a red cylinder, which is stored in the $(1,1)$ position of the warehouse and then processed for \SI{1}{s} in the \ac{MPU} before being passed from the sorting station to the delivery bay.
Fig.~\ref{fig:example} shows the respective sensor values (distinguished by line style) over time (measured in process cycles) for each component of the Fischertechnik factory (\textcolor{orange}{orange}) and \name \ (\textcolor{blue}{blue}).
We observe major overlap of all curves, indicating nearly identical behavior of both factories.
Most notably, the shape of all curves is identical and the only deviations result from timings, e.g., the ``shift'' along the x-Axis for the \ac{VC}'s (Fig.~\ref{fig:example} upper right).
These results indicate that \name{} mimics the physical reference factory accurately.

\begin{figure}
  \includegraphics[width=\columnwidth]{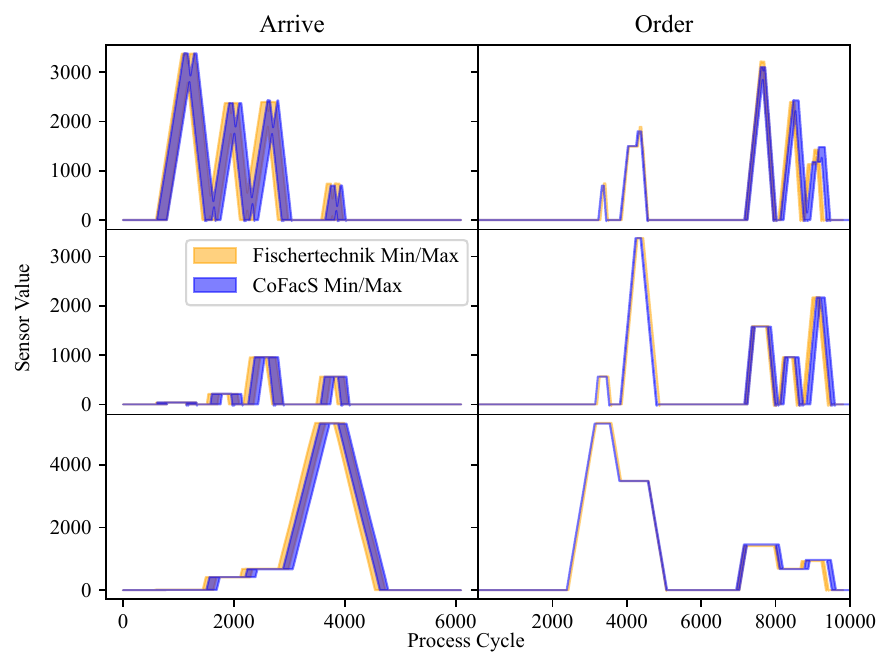}
  \caption{To validate the behavior of \name{}, we compare the \ac{VC}'s movement in the \emph{arrive} (left) and \emph{order} (right) scenario. We show the minimum and maximum value for the Fischertechnik Learning Factory 4.0 (\textcolor{orange}{orange}) and \name{} (\textcolor{blue}{blue}) and the \emph{vertical} (top), \emph{horizontal} (middle), and \emph{rotation} (bottom) sensor. We observe a maximum deviation of 0.11~\% for all sensors, demonstrating the accuracy of \name{}.}
  \label{fig:validation}
\end{figure}

\noindent
\textbf{Detailed Validation.}
To further illustrate that \name{} accurately represents the Fischertechnik factory, we comprehensively analyze the behavior of the \ac{VC} as the most complex component.
We measure the \ac{VC}'s vertical (Fig.~\ref{fig:validation} top) and horizontal (Fig.~\ref{fig:validation} middle) position and the rotational angle (Fig.~\ref{fig:validation} bottom) during four runs of the physical reference factory and compare them to four simulation runs of \name{}.
We divide our measurements into two scenarios: \emph{arrive}, where the \ac{VC} transports a cylinder from material delivery to the warehouse, and \emph{order}, where the \ac{VC} transports a cylinder from the warehouse to the \ac{MPU}.
Fig.~\ref{fig:validation} visualizes the minimum and maximum values of the sensors for the physical (\textcolor{orange}{orange}) and virtual (\textcolor{blue}{blue}) factory.
Again, we observe a substantial overlap of the sensor values.
To examine the maximum deviation of the individual sensors' value, corresponding to a deviating movement of the \ac{VC}, we measure the relative distance between the curves from the physical Fischertechnik factory and \name{}.
Beginning with the vertical sensor, we observe a relative deviation of \emph{\SI{0.00}{\percent}} in the arrive and \emph{\SI{0.05}{\percent}} in the order scenario.
The horizontal sensor exhibits a maximum deviation of \emph{\SI{0.10}{\percent}} and \emph{\SI{0.03}{\percent}}, and the rotation sensor a maximum deviation of \emph{\SI{0.11}{\percent}} and \emph{\SI{0.03}{\percent}} for the respective scenarios.
These results show that \name{} replicates the behavior of the physical reference factory \emph{very} closely.
Therefore, \name{} serves as an accurate virtual testbed to study attacks without risking damage to real components and provides further insight into the security of interconnected production.

\noindent
\textbf{Resource Requirements.}
By performing all experiments on a Raspberry Pi 5, we already show the rather low resource requirements of \name{}.
During the execution of the simulation, we measure approximately \SI{40}{\percent} CPU usage and \SI{1}{GB} of memory usage.
These modest requirements result not only in \textit{scalability} of \name \ (e.g., by being able to simulate more components on more powerful hardware), but they also provide high \textit{accessibility}, since low-cost hardware suffices to run \name{}.

\begin{takeaway}
Our results show that \name \ achieves a maximum deviation of \SI{0.11}{\percent} from the physical reference factory, thus providing a realistic representation of a complex factory environment.
As these results were obtained on rather low-end hardware, \name{} positions itself as an efficient, flexible, accessible and thus widely applicable testbed.
\end{takeaway}

\section{Analysis of Exemplary Attacks on \name{}}
\label{sec:attack}

After validating \name{}, we use it to conduct exemplary attacks on the \textit{physical process} to demonstrate that our testbed accurately reflects behavior under anomalous conditions.
Additionally, we carry out \textit{network} attacks that cannot be performed on the physical testbed due to the risk of damaging expensive components.
These attacks are intended to showcase the versatility of \name{} as a security testbed.
Particularly, by simulating a complete factory, \name{} enables the study of attack impact on multiple production components.
\begin{figure}
  \includegraphics[width=\columnwidth]{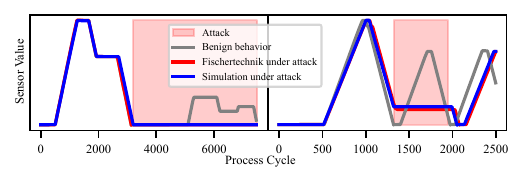}
  \caption{To validate that \name{} accurately replicates anomalous behavior from the physical reference, we perform two low-impact physical attacks on both testbeds. To this end, we remove a cylinder during the execution of the \ac{MPU} (left) and block the vacuum gripper while transporting a cylinder (right).}
  \label{fig:attacks}
\end{figure}

\noindent\textbf{Physical Attacks.}
In a first step, we perform two low-impact physical attacks (i.e., attacks that do not damage the equipment) on the Fischertechnik factory \emph{and} \name{} to verify that our testbed behaves similar to the real system also under anomalous process states.
As a first attack, we \emph{remove the cylinder} after it has been processed by the furnace.
We realize this attack by physically picking up the cylinder from the delivery bay of the \ac{MPU} (physical testbed) and by deleting the corresponding file which simulates a cylinder (\name{}).
In Fig.~\ref{fig:attacks} (left), we depict the sensor readings of the \ac{MPU}'s transport system.
Under normal behavior (\textcolor{gray}{gray}) the \ac{MPU} transports the cylinder from the furnace to the mill, and then pushes the finished product onto the sorting station's conveyor belt, visualized by the two peaks in Fig.~\ref{fig:attacks}. 
In contrast, if the attacker physically removes the cylinder, it cannot be transferred to the sorting station after completing the milling step, visualized by the transport system remaining at $0$ after the initial peak.
Both the physical testbed (\textcolor{red}{red}) and \name{} (\textcolor{blue}{blue}) exhibit identical behavior.

Similarly, for the second attack, we \emph{physically block the vacuum gripper}, which results in the \ac{VC} halting and then resuming movement after being released. 
Fig.~\ref{fig:attacks} (right) shows the reading of the rotation angle of the \ac{VC}, and we again observe nearly identical behavior for the Fischertechnik factory (\textcolor{red}{red}) and \name{} (\textcolor{blue}{blue}), which is clearly distinct from normal behavior (\textcolor{gray}{gray}). 
The only deviation between the behavior of both testbeds stems from halting the \ac{VC} at different process cycles in the physical reference model.

\begin{figure}
  \includegraphics[width=\columnwidth]{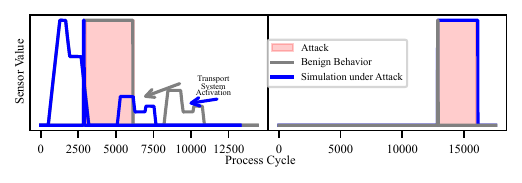}
  \caption{
    By performing two command injection attacks on both components of the \ac{MPU} (i.e., furnace (left) and mill (right)), we verify that \name{} behaves as expected in scenarios we cannot study in a real setup.
  }
  \label{fig:command_injection}
\end{figure}

\noindent
\textbf{Network Attacks.}
Making use of \name{} as a virtual testbed, we execute two \textit{command injection} attacks on the components of the \ac{MPU} which we could not launch against the reference testbed.
In this attack, we model a compromised PLC enabling the attacker to change the system behavior at will.
To cause maximum damage to the production, the attacker chooses to increase the activation times of furnace and mill.
In Fig.~\ref{fig:command_injection} (left), we see the impact of this attack through the increased firing time of the furnace.
Furthermore, this attack causes a delay in the activation of the \ac{MPU}'s transport system.
Similarly, we observe the increased processing time of the mill (Fig.~\ref{fig:command_injection} right) compared to \name{}'s benign behavior.
Additionally, this attack causes a delayed activation of the piston which pushes the ``finished'' cylinder to the sorting station.
In a real system, these attacks could have drastic consequences such as the destruction of materials or even potential fires.

By conducting these attacks, we demonstrate that \name{} enables studying handcrafted protocol-level attacks on specific factory components, posing a significant threat to real production systems~\cite{knapp2024_CH10}.

\noindent
\textbf{Utilizing Existing Attack Tools.}
To enable the study of security measures against known threats, \name{} also provides the possibility to execute existing attack tools, such as \textit{nmap} or \textit{Metasploit}~\cite{metasploit}.
For example, we can launch attacks against the \ac{SCADA} simulation by utilizing the (currently 331) HTTP exploits present in the Metasploit library  over the host machine's localhost interface.
These allow us, e.g., to spy for unencrypted login credentials or active scanning attacks, which often serve as the first step when attacking \acp{ICS}.
Therefore, \name{} can also promote research of security measures against known attacks, further demonstrating its versatility.
\newpage

\vspace*{0.00in}
\begin{takeaway}
  Not only does \name{} precisely replicate the behavior of the physical reference testbed also under attack conditions, but it also enables us to study the impact of attacks that cannot be performed against the real testbed, showcasing \name{}'s versatility. 
  Additionally, \name{} enables the execution of existing security tools such as Metasploit.
\end{takeaway}

\section{Security Research Enabled by \name{}}
\label{sec:intrusion}

To further demonstrate how \name{} enables security research of interconnected production, we conduct case studies on intrusion detection (Sec.~\ref{sec:ids-study}) and the resilience of a 5G-enabled factory against jamming attacks (Sec.~\ref{sec:use-cases:5g}).
Hence, we not only show \name{}'s broad applicability but also lay the foundation for further research ideas that our testbed enables.

\subsection{Intrusion Detection in Interconnected Production}
\label{sec:ids-study}
\acfp{IDS} allow for timely detection of ongoing attacks~\cite{wolsing2022ipal,wolsing2025geco}.
As threats against modern \acp{ICS} not only target the communication but also the physical process (cf.~Sec.\ref{sec:attack}), it is crucial to deploy mechanisms to detect anomalies in both dimensions.
To show how \name{} can facilitate such research, we examine the performance of \acp{IDS} in threat scenarios targeting these dimensions.

\noindent
\textbf{Experimental Setup.}
To conduct these experiments, we create a dataset consisting of benign behavior and attack data.
We capture the benign behavior (i.e., training data for anomaly-based \acp{IDS}) by recording \name{}'s network traffic using Wireshark.
When gathering the attack data (i.e., a command injection attack and a Modbus scan attack), we label all attack timings to generate an accurate ground truth.
We transcribe the Modbus traffic from \name{} into the protocol-agnostic \textit{IPAL} framework~\cite{wolsing2022ipal} to evaluate various state-of-the-art \acp{IDS} on the dataset.
More specifically, we deploy two of the \emph{process-aware} SIMPLE detectors~\cite{wolsing2022simple}, which monitor minimum and maximum or consistent timings of process states, as well as the two \emph{communication-based} \acp{IDS} \ac{IAT}~\cite{lin2018iat}, which monitors timing consistency of network packets, and DTMC\cite{ferling2018DTMC}, monitoring packet sequences.

\begin{figure}
  \includegraphics[width=\columnwidth]{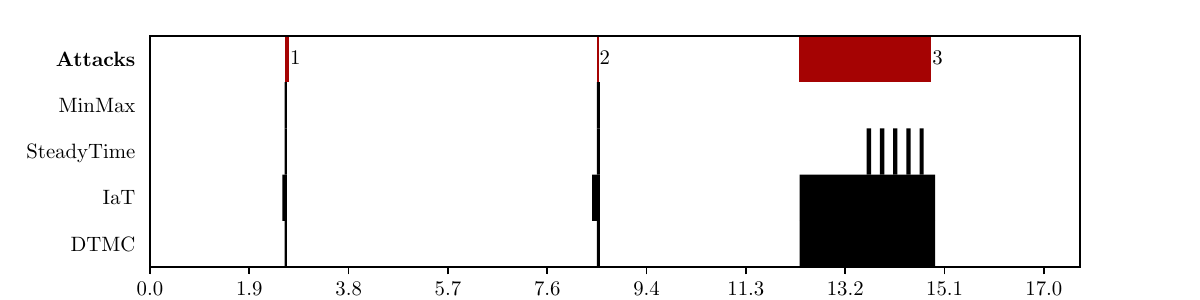}
  \caption{
    The process-aware \acp{IDS} \textit{MinMax} and \textit{SteadyTime}~\cite{wolsing2022simple}, along with the communication-based \acp{IDS} \textit{IaT}~\cite{lin2018iat} and \textit{DTMC}~\cite{ferling2018DTMC}, are capable of detecting  command injection attacks (1 \& 2) against \name{}. However, not all \acp{IDS} reliably detect the Modbus-Scan Attack (3) highlighting the need to capture multiple dimensions of modern production systems which \name{} enables.
  }
  \label{fig:alerts}

\end{figure}

\noindent
\textbf{Results.}
Figure~\ref{fig:alerts} visualizes the alerts of the \acp{IDS} in reference to the ground truth (top, red bars).
All \acp{IDS} can detect the injection attacks (Fig.~\ref{fig:alerts}, 1 \& 2), as the additional packet not only causes a change in the process state (MinMax and SteadyTime), but also sufficient deviation in the network patterns for \ac{IAT} and DTMC to notice the attack.
As expected, the Modbus Scan attack (Fig.~\ref{fig:alerts}, 3) impacts communication and both communication-based \acp{IDS} (\ac{IAT} and DTMC) detect this attack. 
In contrast, \acp{IDS} monitoring whether physical process variable stay within bounds (MinMax) cannot detect such an attack.
Interestingly, however, while also only focussing on process variables, SteadyTime still detects (albeit delayed) that monitored process registers are updated later due to the attack.
These results demonstrate how a complete factory simulation can provide novel insights into the performance of \acp{IDS}. 
Consequently, this case study highlights the need for a testbed that accurately captures all aspects of a modern factory to adequately study \ac{IDS} approaches.

\subsection{Resilience of 5G-based Factories Against Jamming Attacks}
\label{sec:use-cases:5g}

To match the increasing demands for flexibility and mobility for modern production systems, analyzing the capability of wireless communication in industrial setting is essential~\cite{michaelides2025industry5G,michaelides2025testing}.
In this regard, we show how \name{} can enable research on the resilience of 5G-based communication to attacks on the wireless medium in modern factories.
To this end, we leverage the adaptability of \name{} to replace single links in the emulated factory network with \emph{real 5G communication}.

\begin{figure}
  \includegraphics{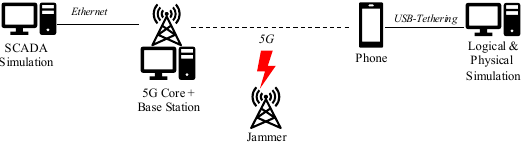}
  \caption{
      We demonstrate \name{} ability to enable security research beyond virtual simulation by deploying a real 5G wireless communication between \ac{SCADA} (left) and logical simulation (right). We utilize this setup to study the impact of jamming attacks against industrial 5G communication. 
  }
  \label{fig:5g_deployment}
\end{figure}

\noindent\textbf{Experimental Setup.}
We deploy the simulation of the \ac{SCADA} system and the control logic simulation on separate (physical) hosts (Fig.~\ref{fig:5g_deployment}). 
Another host runs the 5G system using Open5GS~\cite{open5gs} as the core network and srsRAN~\cite{srsran} as the radio access network.
Connected to this host is an Ettus USRP B210 \ac{SDR} that serves as the antenna of the 5G base station (i.e., the cell tower of a 5G network).
To connect the logical and physical simulation to the 5G network, we use a commercial-off-the-shelf smartphone as the \ac{UE}. 
Finally, a second Ettus USRP B210 \ac{SDR} serves as the attacker performing the \textit{jamming attack} using the CleverJam~\cite{cleverjam} framework.%

\noindent\textbf{Results.}
We verify the successful deployment of \name{} over 5G by observing the behavior of the production process. 
Similar to the fully virtual deployment, we can place new cylinders and observe the expected behavior from the \acp{PLC} and physical simulation when ordering a cylinder through the \ac{SCADA} interface.
However, when activating the jammer, we observe that no further communication via the 5G link is possible.
Thus, the communication between \ac{SCADA} and \acp{PLC} is disrupted rendering monitoring, ordering of goods and adapting the production process (e.g., the firing time) impossible.
Thus, there is a need to develop and evaluate resilience measures for such attacks, for which \name{} provides the ideal basis.
\begin{takeaway}
  As demonstrated by two exemplary case studies, \name{} enables a wide range of security research by comprehensively covering physical and networking behavior.
  Additionally, \name{} allows producing artifacts for detailed studies and is adaptable to also incorporate real components such as an actual 5G network.
\end{takeaway}

\section{Conclusion}

In this paper, we present \name, the first complete factory simulation of an \textit{entire} production process to facilitate security research for interconnected production and \ac{SCADA} applications in an end-to-end manner. %
By comprehensively simulating the physical processes, PLC logic, network communication, and SCADA application of a complete production line, and also precisely emulating the industrial network, \name{} enables the execution of real applications and especially attack tools.
Through comparison with a physical reference, the Fischertechnik Learning Factory 4.0~\cite{FTfactory}, we show that \name{} accurately captures the behavior of all components of the physical production line both under normal conditions and attack situations.
Consequently, we can apply \name{} to perform attacks such as malicious command injection, which cannot be performed against the physical testbed due to the risk of inflicting permanent physical damage.
Additionally, we can extract data to evaluate security mechanisms in detail such as \acp{IDS}.
Furthermore, the flexibility of our testbed allows us to exchange the emulated network with physical components enabling us to study the resilience of 5G communication in production systems against jamming attacks.

To enable follow-up work and thus spark further research to strengthen the security of interconnected production, we make \name{} freely and openly available to the research community~\cite{cofacs}.
Therefore, \name{} creates an ideal environment to evaluate research on industrial networks such as efficient transport security for industrial communication~\cite{bodenhausen2025caching}, classifying encrypted traffic~\cite{encryptedTrafficLCN2023}, object security for industrial data~\cite{henze2013maintaining}, software defined networking in industrial settings~\cite{sdnLCN2021}, or scheduling algorithms for federated learning~\cite{federatedLCN2021}.
As such, with \name{}, we lay the foundation for comprehensively studying the security of modern interconnected factories and thus ultimately improve their security.

\section*{Acknowledgments}

Funded by the Deutsche Forschungsgemeinschaft (DFG, German Research Foundation) under Germany's Excellence Strategy -- EXC-2023 Internet of Production -- 390621612 and the Foundation for Innovation in Higher Education (Stiftung Innovation in der Hochschullehre), Freiraum project RealistICS.

\end{document}